\documentstyle[aps,prb,graphicx,amsmath,array]{revtex}

\newcommand\ud      {{\mathrm{d}}}
\newcommand{\dpa}[2]    {\frac{\partial{#1}}{\partial{#2}}}
\newcommand{\eps}{\epsilon}

\newcommand{\FIG}[6]  {\begin{figure} [#2] \begin{center} 
                        \includegraphics[width=#3cm, height=#4cm]
                        {\fig{#1}}
                        \caption{#6}    \label{#5}  
                        \end{center}    \end{figure}}

\newcommand{\fig}[1]  {./#1}

\begin{document}

\title{Phase transition in the ground state of a particle in a double-well
potential and a magnetic field}
   
\author{H. Kunz$^1$, J. Chapuis$^1$, A. Jory$^2$}

\address{$^1$ Institute of Theoretical Physics, Swiss Federal Institute 
of Technology Lausanne, CH 1015 Lausanne EPFL}
\address{$^2$ Department of Mathematics, Swiss Federal Institute 
of Technology Lausanne, CH 1015 Lausanne EPFL}

\date{\today}

\twocolumn[\hsize\textwidth\columnwidth\hsize\csname @twocolumnfalse\endcsname

\maketitle

\begin{abstract}
We analyse the ground state of a particle in a double-well potential, with a
cylindrical symmetry, and in the presence of a magnetic field. We find that
the azimuthal quantum number $m$ takes the values $m = 0,1,2\dots$ when we increase
the magnetic field. At critical values of the magnetic field, the ground state
is twice degenerate. The magnetisation shows an oscillatory behaviour and jumps
at critical values of the magnetic field. This phase transition could be seen in
the condensate of a dilute gas of charged bosons confined by a double-well
potential.
\end{abstract}

\bigskip

\bigskip

\pacs{03.75 Fc, 03.65 Ge, 31.15 -p}
]

%
\section{Introduction}
It is a well-known fact \cite{RS} that in the absence of a magnetic field the ground
state of bosons is non degenerate, and therefore has the symmetry of the
hamiltonian. Mathematically this result from the fact that the kernel  of the operator
$e^{-tH}$ is positive. This last property no more holds in the presence of a
magnetic field so that degeneracy of the ground state may be expected, as well as
symmetry breaking in it. One-body systems already may show this phenomenon.
Indeed Lavine and O'Carrol \cite{LO} proved the existence of spherically symmetric potentials
for which, in the presence of a magnetic field, the ground state has a
non-vanishing value for the $z$ component of angular momentum, so that the
rotational symmetry is broken. 

Further examples were provided by  Avron, Herbst and  Simon \cite{AHS}$^{,}$
\cite{AHS78}. On the
opposite side, these last authors were able to prove that for the hydrogen atom
the symmetry is not broken, as well as in the case where the potential is
monotonically increasing with the distance. These authors, however, mainly
concerned with problems of atomic physics, did not discuss the degeneracy and
the physical significance of it. 

On the other hand, two of us analysing the problem
of a particle confined to a disc or an annulus in the presence of a magnetic
field found that the ground state was degenerate in the case of an annulus and
for a disc with Neumann boundary conditions (with Dirichtlet boundary conditions
in the disc case the degeneracy disappears.) The degeneracy appears each time the
magnetic field reaches a critical value and the magnetisation jumps at these
critical values, which form a discrete set \cite{jory}.

Motivated by these results we consider in this article a class of systems for
which similar phenomena occur. Namely we analyse the ground state of a particle
in three dimensions moving in a double-well type potential, cylindrically
symmetric, and submitted to a constant magnetic field in the $z$ direction.

We find that the ground state  has an azimuthal momentum $\hbar m$
taking increasing values $m=0,1,2,...$ when we increase the magnetic field $B$.
At critical values of $B$ $(B_m)$ the ground state is twice degenerate between
the $m$ and the $m+1$ state. Moreover the magnetisation jumps at these critical
values and shows in general an oscillatory behaviour reminiscent of the well
known de Haas von Halphen oscillations in solid state physics.

We show that this phenomenon can be understood by an
analysis of the minima of the potential energy, fixing however the angular
momentum to its quantised value $\hbar m$. In the two-dimensional case we can
use the WKB method and obtain bounds on the energy in order to estimate the
critical fields. But in general, we had to compute numerically the energies and
compare them to estimates based on trial wave functions. The agreement is quite
good in general.

Concerning possible experimental verifications of these effects, which require
basically to have a potential which has a minimum sufficiently far from the
origin, we could think of two cases. The first one would be in some molecules
where proton dynamics could be described by such an effective potential. The
second one, more thrilling, would be the case of charged bosons undertaking  a
Bose-Einstein condensation. Our results suggest that in this case, the bosons
would undertake a {\bf phase transition} in their condensate, when we apply an
increasing magnetic field. This phase transition would manifest itself by
appearance of oscillations in the {\bf magnetisation}, which would {\bf jump}
at certain critical values of the magnetic field.


\section{The Model}

We will consider the case of a particle of mass $\mu$, charge $q$, in a potential
$V$ with a cylindrical symmetry, submitted to a magnetic field $\widetilde{B}$ in
the $z$ direction. We do not consider the effect of the spin of the particle.
We choose for a unit of energy $V_0$, and length $r_0$, both being characteristic of
the potential. The dimensionless hamiltonian reads if $r = \sqrt{x^2+y^2}$
	\begin{equation}
	(i \, \eps \, \vec{\nabla} - \vec{A} )^2 + V(r,z)
	\end{equation}
where
	\begin{equation}
	\eps \, = \, \frac{\hbar}{r_0 \sqrt{2 \mu V_0}}	
	\end{equation}
measures the importance of the quantum effects and the vector potential in the
symmetric gauge is given by
	\begin{equation}
	\vec{A} \, = \, \big( \frac{-By}{2},\frac{Bx}{2},0 \big)
	\end{equation}
$B = \dfrac{q}{c} \, \dfrac{r_{0}}{\sqrt{2 \mu V_{0}}} \,  \widetilde{B}$ being the
dimensionless magnetic field.

Thanks to the cylindrical symmetry, we can replace the $z$ component of the
angular momentum $L_z$ by its eigenvalue $\eps m$ so that the reduced
hamiltonian reads 
	\begin{equation}	\label{Hm}
    H_{m}  =  -\epsilon^{2} \left[ \frac{1}{r}\frac{\partial}{\partial r}
	  r \frac{\partial}{\partial r} 
	+ \frac{\partial^{2}}{\partial z^{2}} \right]
	+ \left(\frac{\eps m}{r} - \frac{rB}{2}\right)^{2}+V(r,z) 
	\end{equation}
The ground state energy of this hamiltonian and the corresponding eigenfunction
will be denoted $E_m$ and $\psi_m$.

It remains to specify $V$. We will basically consider a double-well potential of
the form:
	\begin{equation}
 	V(r,z) \, = \, r^4+z^4-2 (r^2+z^2)+v \, r^2z^2
	\end{equation}
with $v$ satisfying $v \, \geq \, -2$, so that $V$ is bounded from
below. If $v$ is equal to $0$ we can decouple the motion in the $z$ direction
form the one in the plane perpendicular to the magnetic field. This is what we
will call the {\bf two-dimensional case}. 
If $v = 2$, we have in three dimensions a potential with spherical symmetry.

We have chosen this double-well form because if we had taken the simple well $V
\, = \, r^4+z^4+2(r^2+z^2)+v \, r^2z^2$ with $v \geq 0$ it follows from \cite{AHS}
that the ground state is not degenerate and
corresponds to $m = 0$.

A physical quantity of interest is the {\bf magnetisation} in the
ground state
	\begin{equation}
 	M \, = \, -\dpa{E}{B}
	\end{equation}
in units $\frac{q}{c} \, r_0  \sqrt{\frac{V_0}{2 \mu}}$

We will denote by $e_m$ the ground state energy of the hamiltonian
	\begin{equation}
    h_{m} \, = \, -\epsilon^{2} \left[ \frac{1}{r}\frac{\partial}{\partial r}
	  r \frac{\partial}{\partial r} 
	+ \frac{\partial^{2}}{\partial z^{2}} \right] +V_m(r,z) 
	\end{equation}
with
	\begin{equation}
    V_m \, = \, \frac{(\eps  m)^2}{r^2} + \frac{B^2}{4}r^2 + V
	\end{equation}
and by
	\begin{equation}
    E_m \, = \, e_m - \eps m  B
	\end{equation}
the ground state energy of $H_m$ given in \eqref{Hm}, so that the real
ground state energy is given by
	\begin{equation}
    E \, = \, \inf_{m \geq 0} E_m
	\end{equation}
since obviously negative $m$ give a larger energy.

Finally we will use the following useful scaling property of the energy
$e_m$
	\begin{equation} \label{rescal}
    e_m(\eps,\lambda,v) \, = \, s^2 \, e_m \Big(\frac{\eps}{s^{ 3/2}},
	\frac{\lambda}{s},v \Big) \qquad \quad \forall \: \: s \geq 0
	\end{equation}
where
	\begin{equation}
    \lambda \, = \, \frac{B^2}{4} -2 
	\end{equation}
is the parameter multiplying $r^2$ in the potential. Equation 
\eqref{rescal} follows simply from the scaling transformation : 
$r^2 \to s r^2$ and $z^2 \to s z^2$. This relation shows that we have
effectively a two parameter dependence of the energy $e_m$ in general and
a one parameter dependence in the two dimensional case.

The choice $s = |\lambda|$ or $s = m^{\frac{1}{3}} \; (m \geq 1)$ shows that
large magnetic field or large angular momenta correspond to the
semi-classical limit.
In fact we shall see that in the classical limit $\eps \to 0$ ground state
with $m \neq 0$ are favoured inducing ground state degeneracies at some
values of the magnetic field. It thus appears that the tendency to have a
ground state with the same symmetry as the hamiltonian and therefore non
degenerate is an effect due to quantum mechanics.


\section{The classical limit}

One can gain some qualitative understanding of the problem by looking at the
classical limit of it. This means that we neglect the quantum kinetic energy and
define the ground state energy as 
	\begin{equation}
    E \, = \, \inf_{m \geq 0} \: \inf_{(r,z)} \:[V_m(r,z) - \eps m B]
	\end{equation}
where
	\begin{equation}
    V_m \, = \, \frac{(\eps m)^2}{r^2} + r^4 + z^4 + \frac{B^2}{4}r^2 
	-2r^2 +v\, r^2 z^2 
	\end{equation}
and consider that $m$ is an integer.

Two cases need to be considered separately: $|v| < 2$ and $v \geq 2$. 
If $|v| < 2$ we denote by $x_m$ and $t_m$ respectively the values of
$r^2$ and $z^2$ which minimise the potential $V_m$, and we find
	\begin{gather}	\label{tmxm}
	t_m \, = \, 1 - \frac{v \, x_m}{2} \\ \nonumber
	\left(2-\frac{v^2}{2} \right) x_m + \left( v-2 + \frac{B^2}{4} \right) = 
	\frac{(\eps m)^2}{x^2_m}	
	\end{gather}
On the other hand, considering for a while $m$ as a continuous variable, the
absolute minimum of $V_m - \eps m B$ is given by
	\begin{equation}
    \eps \hat{m} \, = \, \frac{B}{2}x_{\hat{m}}
	\end{equation}
From \eqref{tmxm} this gives an
absolute minimum of $V_m - \eps m B$  given by
	\begin{equation}
    x_{\hat{m}} \, = \, t_{\hat{m}} \, = \, \frac{1}{1 + \frac{v}{2}}
	\end{equation}
and therefore
	\begin{equation}
    \eps \hat{m} \, = \, \frac{B}{2} \frac{1}{1 + \frac{v}{2}}
	\end{equation}
In considering the variable $m$ as a continuous one we have treated the problem
purely classically and the corresponding ''ground state'' energy is
	\begin{equation}
    E^{cl} \, = \, - \frac{2}{1 + \frac{v}{2}}
	\end{equation}
We know that $m$ is a discrete variable but for consistency we must consider
$\eps$ as a small number. Then if $m$ designates the integer part of $\hat{m}$,
we have $\hat{m} = m + \theta$ and if $0 \leq \theta < \frac{1}{2}$, the ground
state has the quantum number $m$, whereas if $\frac{1}{2} < \theta \leq 1$ it
has $m + 1$.

From this analysis we conclude that if $B_{m-1} < B < B_m$ where
	\begin{equation} \label{Bm}
    B_m \, = \, \eps (1+\frac{v}{2})(2m+1)
	\end{equation}
the ground state has the quantum number $m$.  Hence we see that by increasing the
magnetic field, we find in increasing order the values of $m = 0,1,2,...$ and an
infinite set of  {\bf critical values of the magnetic field exist}, $B_m$ for which
the ground state is twice  {\bf degenerate}, being both $m$ and $m+1$. \\ This
picture is entirely confirmed by the numerical results in the quantum case. It is
also quite interesting to look at the magnetisation. In the state whose
quantum number is $m$, we have 
	\begin{equation}
    M_m \, = \, \eps m - \frac{B}{2}x_m
	\end{equation}
so that using \eqref{tmxm}
	\begin{equation}	\label{jmpeq}
    M_m \, = \, [\eps m - \frac{B}{2} \frac{1}{1+\frac{v}{2}}]
	[\frac{1-\frac{v}{2}}{1-\frac{v}{2} + \frac{B^2}{4}}]
	\end{equation}
when $B_{m-1} < B < B_m$. 

This shows that the magnetisation has an ''oscillatory'' type of behaviour
reminiscent of the familiar de Haas von Halphen one in solid state physics and
that the {\bf magnetisation jumps at the critical values of the
magnetic field}, the jump being given by
	\begin{equation}
    \Delta M_m \, = \, \eps \frac{1-\frac{v}{2}}{1-\frac{v}{2}+\frac{B^2}{4}}
	\end{equation}

Once again this general behaviour is reproduced by the numerical results in the
quantum case and the spacing between the values of the critical field is rather well
represented by formula \eqref{Bm} when $m \geq 1$. In the 
{\bf two-dimensional} case, i.e. $v = 0$ and neglecting the trivial $z$
dependence, we can proceed further and look at a really semi-classical approximation
namely WKB, for the ground state energy
	\begin{equation}
    \int_{r_-}^{r_+} \ud r \sqrt{e_m - V_m(r)} \, = \, \frac{\eps \, \pi}{2}
	\end{equation}
where
	\begin{equation}
    V_m(r) \, = \, \frac{(\eps m)^2}{r^2} + r^4 +  
	\big(-2 + \frac{B^2}{4} \big)r^2 
	\end{equation}
and the ground state energy is $E_m = e_m - \eps m B$.

In fact this WKB approximation will give the best analytical results, apart from
the variational estimates for the energy, which give unfortunately only exact upper
bounds on the energy.

When the potential has spherical symmetry $v = 2$, quantum effects are much more
important and the classical analysis gives only that the ground state has $m =
0$ if $B < 2\eps$, is degenerate between $m = 0$ and $m = 1$ when $2 \eps \leq B 
< 4 \eps$, has possibly $m = 0,1,2$ for $4 \eps \leq B < 6 \eps$ and so on.
This only suggests that we have again the increasing sequence of $m$, when we
increase the magnetic field and that critical values appear near $2 \eps m$.

When $v > 2$, we find that $m = 0$ is the ground state except when $B = 2 \eps
m$, where it is degenerate between $m$ and $0$. We may note however that the
classical ground state correspond to  points $(r=0,z= \pm 1 )$   in configuration
space for $m=0$, whereas it corresponds to two circles $(r=\frac{\eps}{2B}, z = \pm
 \sqrt{1-\frac{\eps}{2B}})$ for $m=1$ and $2\eps < B < 4\eps$, so that the wave
 function can be more spread in the $m=1$ state than the in the $m=0$ state, and that 
the kinetic energy of the $m=1$ state is lower, favouring the $m=1$ state. Hence 
we should expect, at least when $\eps$ is small, a ground state with $m=0$ for 
small fields and a ground state with $m=1$, when  $2\eps < B < 4\eps$. A similar 
argument can be given for the higher values of $m$.

Finally, it is worth noticing that if we had taken a simple well type potential
	\begin{equation}
 	V(r,z) \, = \, r^4+z^4 + 2 (r^2+z^2)+v \, r^2z^2
	\end{equation}
the classical analysis gives a ground state with $m = 0$, at least when $v \geq
-1$. This is a correct result when $v \geq 0$ at the quantum level.
%

%

\newpage

\section{Numerical results and variational bounds}

It is quite useful to undertake a numerical analysis of this problem. 
We have used a finite element method, choosing for the basis a product of two
triangles functions. We discuss separately the two-dimensional problem and the
three dimensional ones.

\subsection{Two dimensions}

\begin{figure} [!ht]
	\begin{center}
	\hspace{-0.3cm}
    \includegraphics[width=8.4cm, height=8.2cm]{\fig{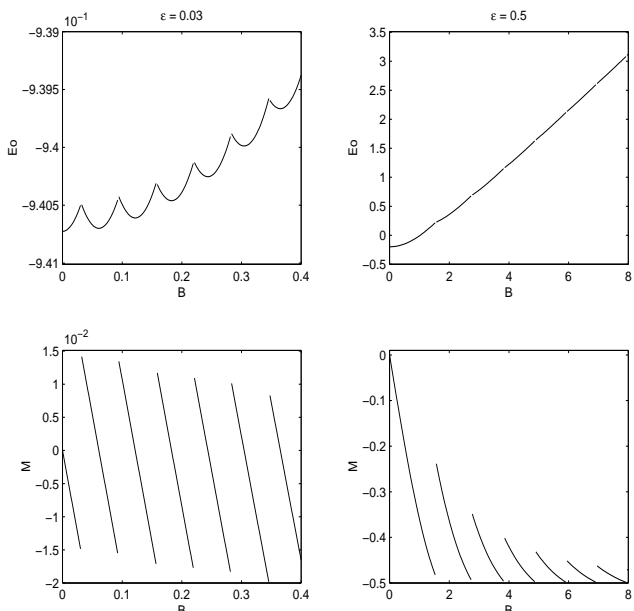}}
    \caption{Energie and magnetisation dependence of $B$
	for $\eps = 0.03$ and $\eps=0.5$}
	\label{c01}  
	\end{center}
    \end{figure}

We first give pictures of the ground state energy for two typical values of
$\eps$, a small ($\eps = 0.03$) and a large one ($\eps = 0.5$) as a function of the
magnetic field $B$. (figure \ref{c01}). The cusps at the critical values of $B$
indicate a jump of the corresponding magnetisation. 

This last quantity shows
first a diamagnetic behaviour at small field, but then a paramagnetic -
diamagnetic oscillation at least when $\eps \lesssim 0.3$. Beyond this value
the magnetisation is entirely negative (figure \ref{c01} bottom right). We can also note
that when $B$ becomes large the magnetisation tends to $-\eps$, its value in
the Landau regime.

The results clearly indicate that we go progressively from the states with $m =
0,1,2 \dots$ by increasing the magnetic field and that the magnetisation jumps at
the critical values. The effect is more pronounced in the classical regime.
All these results are in qualitative argument with the classical picture 
presented before and the agreement is even quantitative when $\eps = 0.03$ for
example.
 	\FIG{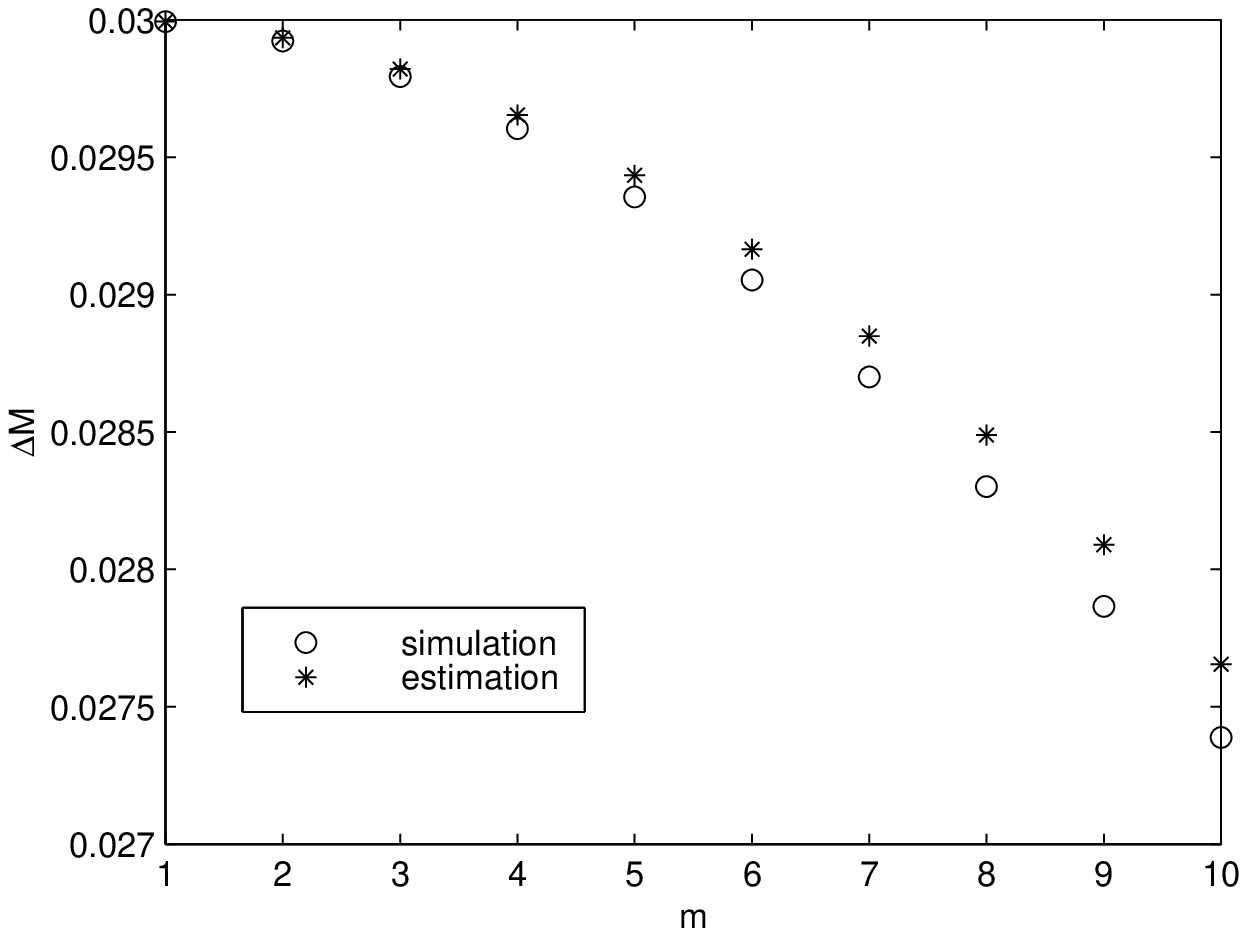}{!ht}{6.5}{5.5}{jmp}{Comparison of the jumps given by
	simulation and the estimation}
The jumps of the magnetisation given by formula \eqref{jmpeq} are reproduced 
(figure \ref{jmp}) with a
precision of less than $1$ percent when $\eps = 0.03$, and the spacing between
the critical values of the magnetic field
	\begin{equation}
    \frac{B_{m+1} - B_m}{\eps} \, = \, 2 + \Delta_m
	\end{equation}
is given by $\Delta_m \leq 0.04$ if $m \geq 1$ and $\eps = 0.1$. $\Delta_m$
decreases when $m$ increases in agreement with the scaling relation 
$B_m = (2m+1)\eps$, so
that the simple classical formula reproduces rather well the results. By
contrast, the jump between the $m=0$ and the $m=1$ state is largely of quantum
mechanical origin, as well as the precise values of the critical fields.

	\FIG{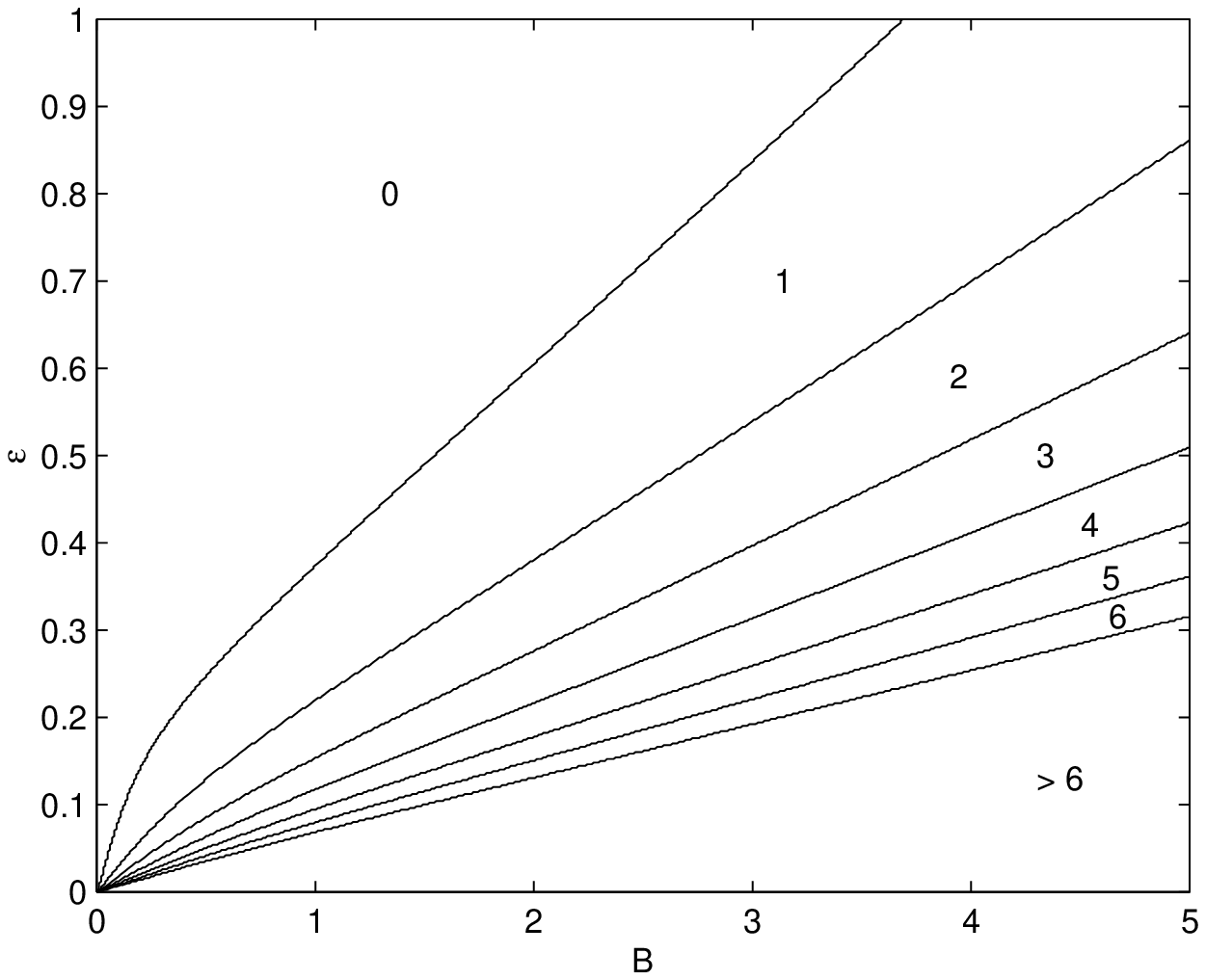}{!ht}{8}{6}{plan2d}{$M$ values of the ground state
	depending on  $B$ and $\eps$}
Figure \ref{plan2d} describes the various regions in the $\eps - B$ plan. We can
note that even when $\eps > 0.25$ a linear relation exists between $B_m$ and
$\eps$, as in the classical regime, which is a bit surprising.
	\begin{figure} [!ht]
	\begin{center}
	\hspace{-0.3cm}
    \includegraphics[width=8.2cm, height=7.5cm]{\fig{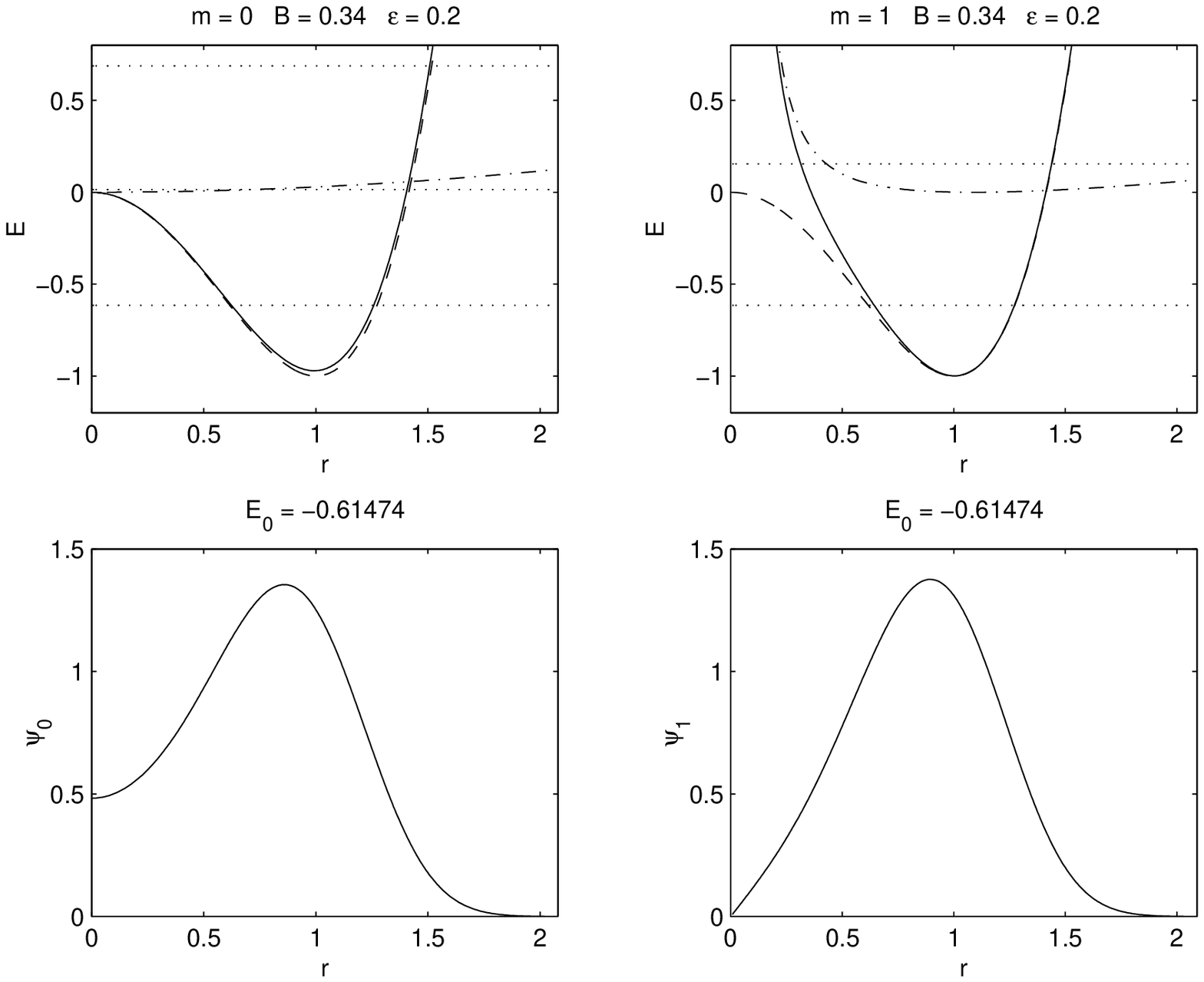}}
    \caption{Potential and eigenfunction of the ground state
	for $m = 0$ (left) and $m=1$ (right) with
	$V_m$ ($-\!-$)  = orbital kinetic energy ($- \cdot -$) + 
	double-well ($-\,-$) and energy levels ($\cdot\cdot\cdot$) }
	\label{fct}  
	\end{center}
    \end{figure}

It is also interesting to look at the eigenfunctions when the magnetic
field reaches its critical value. In figure \ref{fct} we give pictures of them at the
critical value between the state $m = 0$ and $m = 1$ when $\eps = 0.2$. We see
that their maxima are located very near the minimum of the potential.

Finally we compare the results with two theoretical estimates: first of all the
WKB one, and a variational one. This last estimate is based on the following two
parameters trial wave function
	\begin{equation}
    \psi_m \, = \, r^m \,  e^{-\alpha r^2 - \beta (r-1)^2}
	\end{equation}
The variational upper bound on the energy can be expressed in terms of Weber
cylindrical functions, but we directly computed the corresponding integrals.
	\begin{table}	[!ht] 
	\begin{center}
	\begin{tabular}{|c|c|c|c|c|c|}
	\hline	
    Deg. &  Simul.  &  \multicolumn{2}{c|}{ WKB $\hspace{0.2cm} \delta$ \%}   & 
	\multicolumn{2}{c|}{Variat. $\hspace{0.1cm} \delta$ \%}   \\
    \hline\hline
      0-1  &   0.0313   &   0.0314  &	 0.23  &   0.0317  & 1.16 \\
      1-2  &   0.0944   &   0.0942  &	-0.15  &   0.0946  & 0.23 \\
      2-3  &   0.1573   &   0.1571  &	-0.13  &   0.1574  & 0.09 \\
      3-4  &   0.2201   &   0.2198  &	-0.13  &   0.2203  & 0.06 \\
      4-5  &   0.2830   &   0.2826  &	-0.12  &   0.2830  & 0.02 \\
      5-6  &   0.3457   &   0.3453  &	-0.12  &   0.3458  & 0.02 \\
      6-7  &   0.4085   &   0.4080  &	-0.12  &   0.4085  & 0.00 \\
    \hline	
	\end{tabular}
	\normalsize
	\caption{Magnetic field $B_m$ at the seven first degeneracies with $\eps = 0.03$}
	\label{tabBc003}
	\end{center}
	\end{table}
	\begin{table}	[!ht] 
	\begin{center}
	\begin{tabular}{|c|c|c|c|c|c|}
	\hline	
     Deg. &  Simul.  &   \multicolumn{2}{c|}{  WKB $\hspace{0.3cm} \delta$ \% }   & 
	\multicolumn{2}{c|}{Variat. $\hspace{0.2cm} \delta$ \%} \\
     \hline\hline
      0-1  &   -0.9405   & -0.9401 &   0.66  &  -0.9403 &  0.26 \\
      1-2  &   -0.9404   & -0.9400 &   0.65  &  -0.9403 &  0.27 \\
      2-3  &   -0.9403   & -0.9399 &   0.64  &  -0.9401 &  0.28 \\
      3-4  &   -0.9401   & -0.9397 &   0.63  &  -0.9399 &  0.29 \\
      4-5  &   -0.9399   & -0.9395 &   0.61  &  -0.9397 &  0.30 \\
      5-6  &   -0.9396   & -0.9392 &   0.59  &  -0.9394 &  0.30 \\
      6-7  &   -0.9392   & -0.9389 &   0.56  &  -0.9390 &  0.30 \\
    \hline	
	\end{tabular}
	\normalsize
	\caption{Energies $E_m$ at the seven first degeneracies with $\eps = 0.03$}
	\label{tabEc003}
	\end{center}
	\end{table}
Tables  \ref{tabBc003},\ref{tabEc003},\ref{tabBc05} and \ref{tabEc05}
 give a comparison of the results for two values of the
parameter $\eps$, and for the critical fields. Excellent agreement is found for
the variational method (maximal error of the order of 2 \% when $\eps = 0.5$).
WKB works quite well when $\eps$ is small ($\eps = 0.03$) as expected, but even
 better on the energies when $\eps = 0.5$ 
 and the error does not exceed 1\%.
	\begin{table}	[!ht] 
	\begin{center}
	\begin{tabular}{|c|c|c|c|c|c|}
	\hline	
    Deg. &  Simul.  &  \multicolumn{2}{c|}{WKB $\hspace{0.1cm} \delta$ \%}   & 
	\multicolumn{2}{c|}{Variat. $\hspace{0.0cm} \delta$ \%}   \\
    \hline\hline
      0-1  &   1.538   &  1.661 & 7.95   &  1.508 & -1.98  \\
      1-2  &   2.747   &  2.811 & 2.33   &  2.743 & -0.15  \\
      2-3  &   3.842   &  3.882 & 1.06   &  3.842 &  0.02  \\
      3-4  &   4.891   &  4.919 & 0.56   &  4.894 &  0.05  \\
      4-5  &   5.920   &  5.940 & 0.34   &  5.924 &  0.07  \\
      5-6  &   6.941   &  6.954 & 0.18   &  6.943 &  0.02  \\
      6-7  &   7.953   &  7.964 & 0.12   &  7.956 &  0.02  \\
   \hline				  
	\end{tabular}
	\normalsize
	\caption{Magnetic field $B_m$ at the seven first degeneracies with $\eps = 0.5$}
	\label{tabBc05}
	\end{center}
	\end{table}
	\begin{table}	[!ht] 
	\begin{center}
	\begin{tabular}{|c|c|c|c|c|c|}
	\hline	
     Deg. &  Simul.  &   \multicolumn{2}{c|}{  WKB $\hspace{0.2cm} \delta$ \% }   & 
	\multicolumn{2}{c|}{Variat. $\hspace{0.0cm} \delta$ \%} \\
    \hline\hline
      0-1  &   0.220   & 0.232 &   0.97 &  0.227 & 0.55   \\
      1-2  &   0.685   & 0.686 &   0.04 &  0.690 & 0.25   \\
      2-3  &   1.159   & 1.159 &  -0.02 &  1.163 & 0.16   \\
      3-4  &   1.639   & 1.638 &  -0.03 &  1.642 & 0.12   \\
      4-5  &   2.122   & 2.122 &   0.00 &  2.125 & 0.12   \\
      5-6  &   2.609   & 2.608 &  -0.02 &  2.612 & 0.07   \\
      6-7  &   3.098   & 3.098 &   0.00 &  3.101 & 0.07   \\
    \hline	
	\end{tabular}
	\normalsize
	\caption{Energies $E_m$ at the seven first degeneracies with $\eps = 0.5$}
	\label{tabEc05}
	\end{center}
	\end{table}

%

\subsection{Three dimensions}

\begin{figure} [!ht]
	\begin{center}
	\hspace{-0.2cm}
    \includegraphics[width=8.4cm, height=8.2cm]{\fig{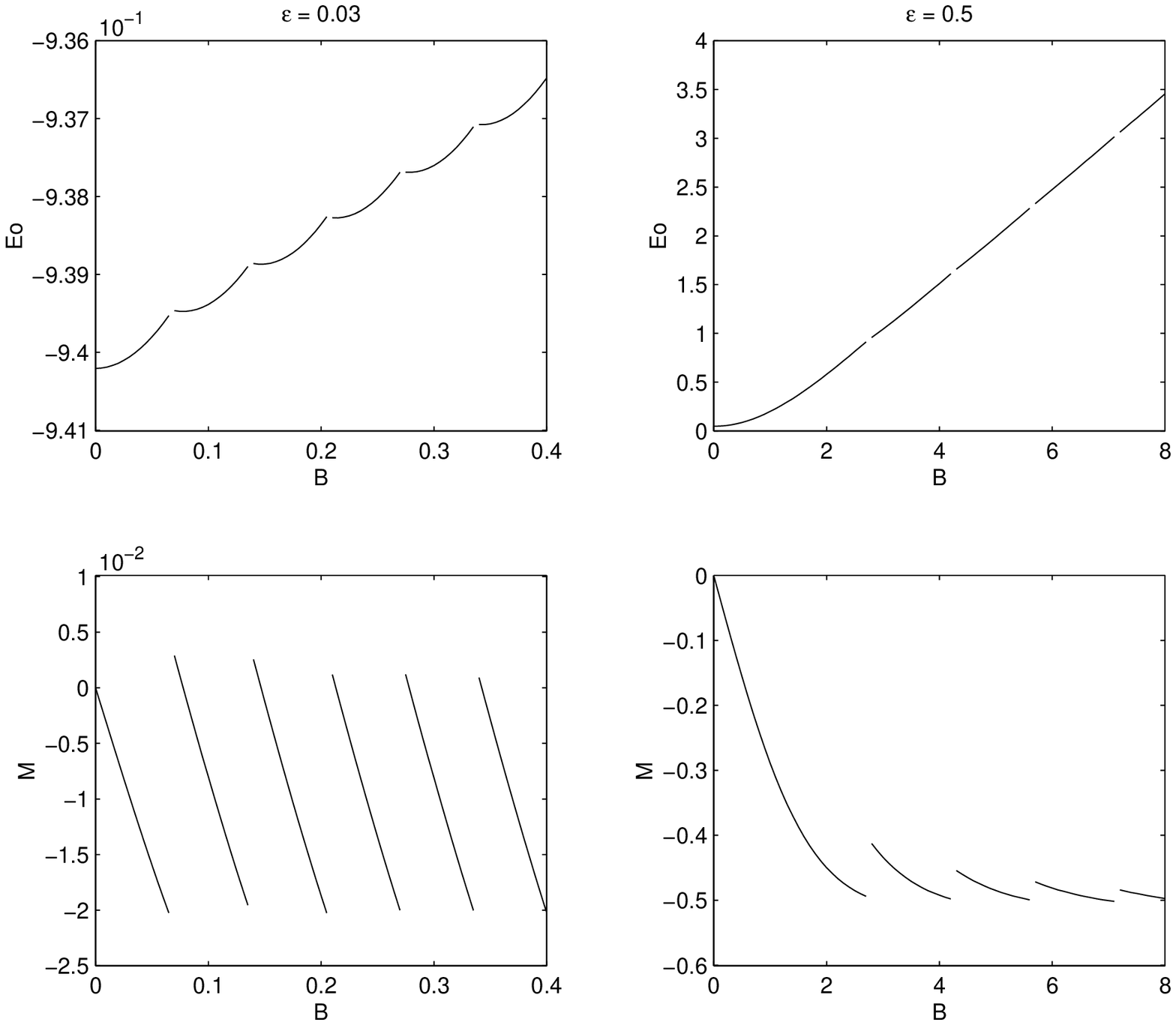}}
    \caption{Energie and magnetisation dependence of $B$
	         for $\eps = 0.03$ and $\eps=0.5$}
	\label{d01}  
	\end{center}
    \end{figure}

For the spherically symmetric potential $(v = 2)$, figure \ref{d01} gives the ground
energies a well as the corresponding magnetisation for two different values of
$\eps$ : 0.03, 0.5.

 Once again we see that the values of $m$ in the ground state
increases with $B$, and that the magnetisation jumps at critical values $B_m$ of
the magnetic field, where the ground state is doubly degenerate.
These results are in qualitative agreement with the classical analysis. Figure 
\ref{plan3d} summaries the results in the $\eps$ - $B$ plane. Notice that in this
  	\FIG{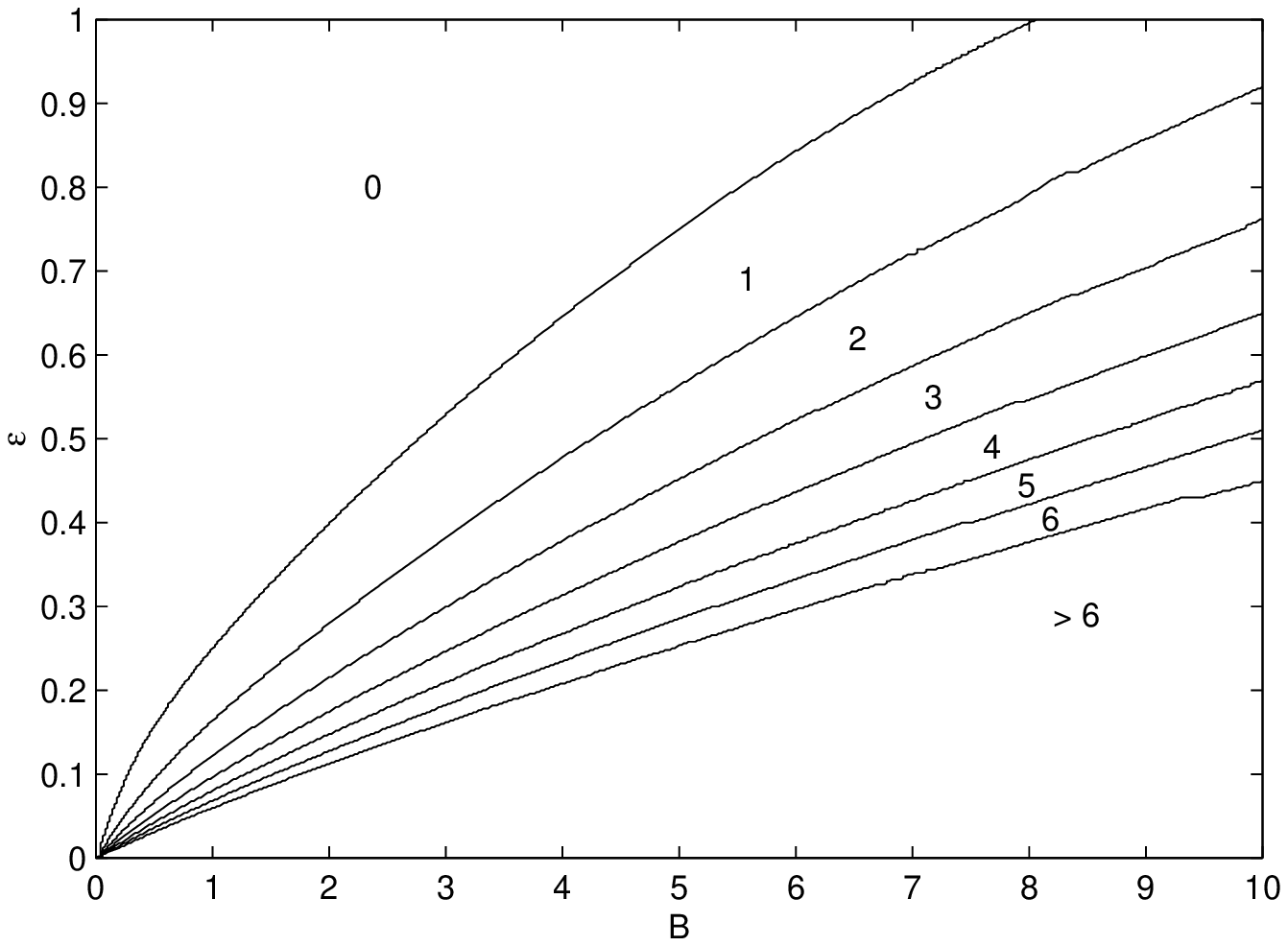}{!ht}{8}{6}{plan3d}{$M$ values of the ground state
  	depending on  $B$ and $\eps$}
case, when $\eps \geq 0.1$ already the relation between $B_m$ and $\eps$ is no more
linear. On the other hand the spacing between the critical values of $B$ predicted
by the crude classical estimate:
	\begin{equation}
        \Delta B_m \, = \, B_{m+1} - B_m \, \cong \, 2\eps
	\end{equation}
is satisfied with a precision of $25\%$ at $m=1$ and becomes more accurate when $m$
increases, at least in the range $\eps \leq 0.1$.

Our best variational estimate for the energy was made with a three parameter trial
wave function 
	\begin{equation}
    	\psi_{\alpha,\beta,\zeta} \, = \, 
	r^m \,  e^{-\alpha r^2 - \beta (\sqrt{r^2+z^2}-\zeta)^2}
	\end{equation}
Table \ref{tab31} gives the values of the critical field $B_m$ and Table \ref{tab32} the
corresponding ground state energies, when $\eps = 0.05$ estimated by the variational
method and computed with the simulation. 
	\begin{table}	[!ht] 
	\begin{center}
	\begin{tabular}{|c|c|c|c|c|c|c|}
	\hline
	 & \multicolumn{2}{c|} {Simulation} & \multicolumn{4}{c|} {Variational} \\
    Deg. & $B_m$ & $E_m$  & $B_m$  &  $\delta_B$ \% &  $E_m$  &  $\delta_E$ \%	  \\
    \hline\hline
      0-1  & 0.1180 &  0.1206 &   2.17  & -0.8986  & -0.8982  &   0.38 \\
      1-2  & 0.2381 &  0.2310 &  -2.94  & -0.8966  & -0.8966  &  -0.01 \\
      2-3  & 0.3549 &  0.3509 &  -1.15  & -0.8946  & -0.8946  &  -0.06 \\
      3-4  & 0.4686 &  0.4616 &  -1.49  & -0.8925  & -0.8925  &  -0.00 \\
      4-5  & 0.5829 &  0.5785 &  -0.74  & -0.8901  & -0.8901  &  -0.01 \\
      5-6  & 0.6961 &  0.6905 &  -0.80  & -0.8876  & -0.8876  &  -0.00 \\
    \hline
	\end{tabular}
	\caption{Magnetic field $B_m$ and energies $E_m$ at the six first 
	degeneracies at $\eps = 0.05$}
	\label{tab31}
	\end{center}
	\end{table}
	\begin{table}	[!ht] 
	\begin{center}
	\begin{tabular}{|c|c|c|c|c|c|c|}
	\hline
	 & \multicolumn{2}{c|} {Simulation} & \multicolumn{4}{c|} {Variational} \\
    Deg. & $B_m$ & $E_m$  & $B_m$  &  $\delta_B$ \% & $E_m$  &  $\delta_E$ \% \\
    \hline\hline
      0-1  &  2.7576	&  0.9415  &  2.6225   &  -4.89  &   0.8959  &  -2.34  \\
      1-2  &  4.2493	&  1.6345  &  4.0912   &  -3.72  &   1.5675  &  -2.54  \\
      2-3  &  5.6746	&  2.3190  &  5.4972   &  -3.12  &   2.2363  &  -2.49  \\
      3-4  &  7.0961	&  3.0126  &  7.0275   &  -0.96  &   2.9845  &  -0.69  \\
      4-5  &  8.5025	&  3.7055  &  8.2415   &  -3.07  &   3.5720  &  -2.83  \\
      5-6  &  9.7537	&  4.3248  &  9.6016   &  -1.55  &   4.2412  &  -1.57  \\
    \hline
	\end{tabular}
	\caption{Magnetic field $B_m$ and energies $E_m$ at the six first 
	degeneracies at $\eps = 0.5$}
	\label{tab32}
	\end{center}
	\end{table}
Obviously there is a very good agreement, since
the largest error for $B_m$ is less than $2\%$ and for $E_m$ less than $0.7\%$.
Table \ref{tab32} gives the same but for $\eps = 0.5$. Again we see a good
agreement (error less than 5\%). 
When $\eps$  increases we found that $\alpha$ increases and $\beta$ decreases
as well as $\zeta$ and our trial wave function becomes less accurate, because
the double-well nature of the potential is less important compared to the
kinetic energy.
\begin{figure} [!ht] 
\begin{center} \hspace{-0.2cm} 
\includegraphics[width=8.5cm, height=6.5cm] {\fig{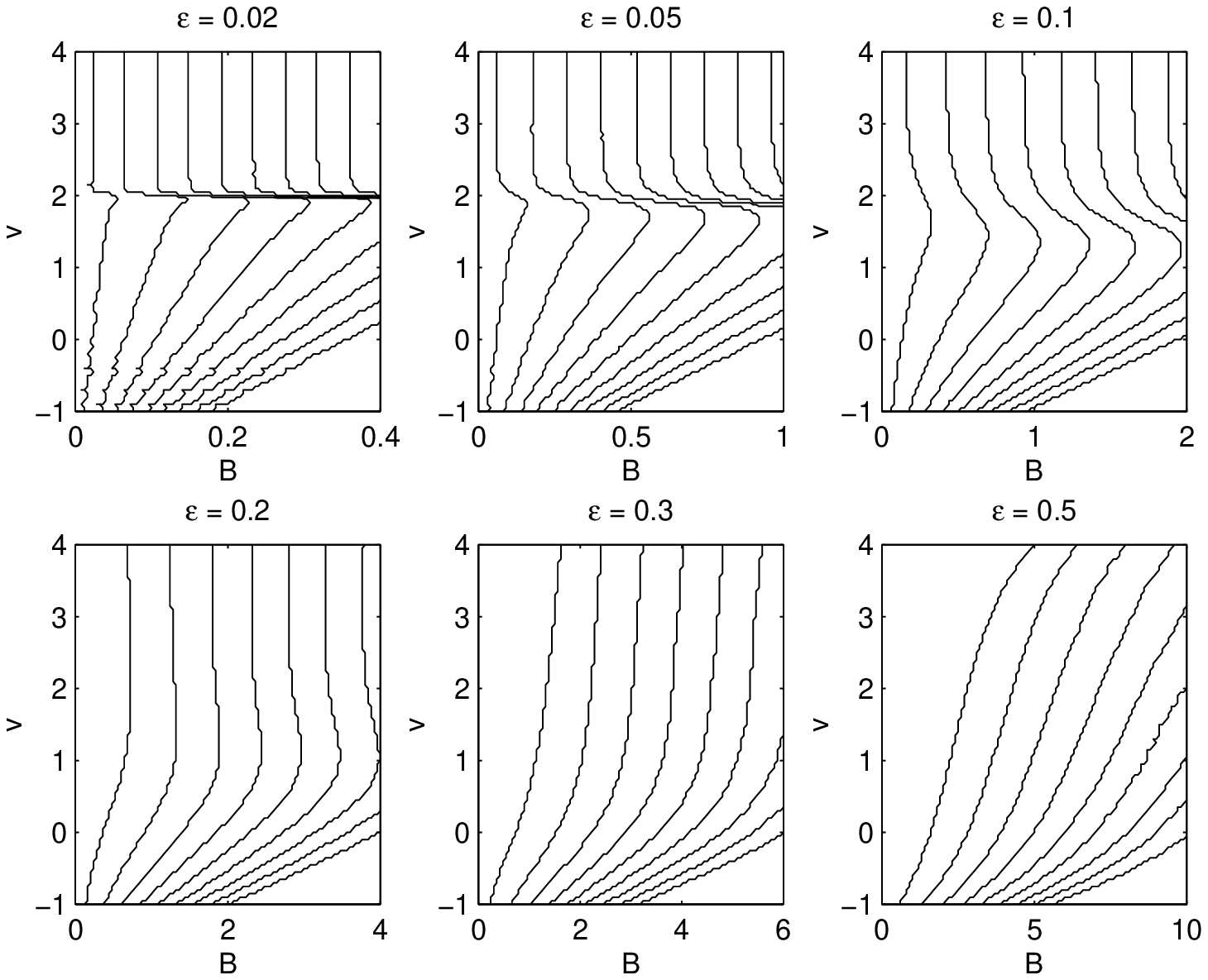}}
\caption{$M$ values of the ground state
  	depending on  $B$ and $v$ with increasing $\eps$}
\label{6gr}  
\end{center}
\end{figure}

Figure \ref{6gr} describes the situation in the $v$ - $B$ plane for $m =
0,1,\dots,10$ and different values of $\eps$. We notice that when $v$ is less than
$2$ and $\eps$ is not too large ($\eps \leq 0.2$), the situation is similar to
the one already discussed, but that there is an abrupt change at $v=2$ when
$\eps$ is small in agreement with the classical analysis. However when  $\eps >
0.2$ the ground state $m=0$ is definitely favoured as $v$ increases.

\begin{figure} [!ht] 
\begin{center} \hspace{-0.1cm} 
\includegraphics[width=8.6cm, height=3.6cm] {\fig{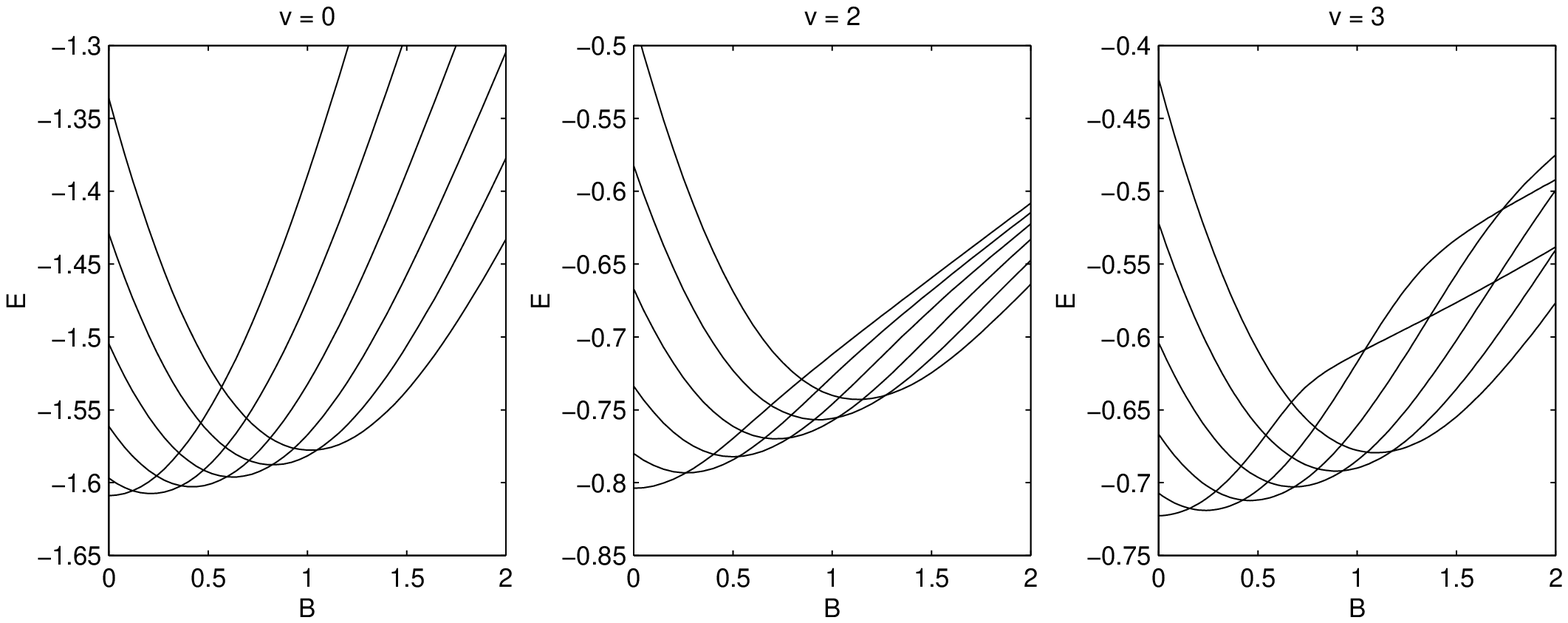}}
\caption{Comparison for different $v$ of the 
	energy for $m = 0,1,2,3,4,5$ depending on $B$ with $\eps = 0.1$}
\label{lines}  
\end{center}
\end{figure}
Figure \ref{lines} shows the energies for the first five $m$ values computed
with three different $v$: two-dimensional ($v = 0$), spherical potential ($v =
2$), and $v = 3$. We can see a new crossing between the $m=0$ and the
other $m$ levels when $v$ becomes larger than $2$, although this does not
concern the ground state.
 

\section{Bounds on the critical field in the two dimensional case}

One might desire to get rigorous upper and lower bounds on the critical fields.
One possible approach would consist in getting upper and lower bounds on the
ground state energies $E_m$. Whereas we have seen that one can obtain very good
variational upper bounds, it is rather difficult to get good lower ones. In
order to test these results, we analysed only the two-dimensional case.

First we want to obtain conditions under which $m=0$ is the ground state. Using
the inequality
	\begin{equation}
    \frac{l^2}{x} + x^2 \, \geq \, x^2 - \frac{l^2}{a^2}x + \frac{2l^2}{a}
	\end{equation}
valid for any $x$ and $a$ positive, we deduce that
	\begin{equation}
    e_0[\lambda] \, \geq \, \frac{2(\eps m)^2}{a} + 
	e_0 \big[\lambda - \left( \tfrac{\eps m}{a} \right)^2 \big]
	\end{equation}
On the other hand
	\begin{gather}
	e_0[\lambda] - e_0 \big[\lambda - \left( \tfrac{\eps m}{a} \right)^2 \big] =
	\int_{\lambda - \left( \frac{\eps m}{a} \right)^2}^{\lambda}
	\ud \lambda' \langle r^2 \rangle_0 (\lambda') \nonumber \\ 
	\leq
	\left(\frac{\eps m}{a}\right)^2 \langle r^2 \rangle_0 
	\big[\lambda - \left( \tfrac{\eps m}{a} \right)^2 \big]
	\end{gather}
since $\langle r^2 \rangle_0 (\lambda)$ is decreasing in $\lambda$.

But 
	\begin{equation} \label{rmoy}
    \big| \langle r^2 \rangle_0 [\lambda] -\frac{\lambda}{2} \big| \, \leq \,
	\Big[e_0 [\lambda] + \left( \frac{\lambda}{2} \right)^2 \Big]^{\frac{1}{2}}
	\end{equation}
The scaling relation and the fact that $e_0$ is increasing in $\eps$ imply that
when $\frac{\lambda}{2} \leq -1$
	\begin{equation}
    e_0 [\lambda] + \left( \frac{\lambda}{2} \right)^2 \, \leq \,
	\left( \frac{\lambda}{2} \right)^2 (e_0[-2] +1)
	\end{equation}
Taking now $a$ such that $\frac{\eps m }{a} \geq \frac{B}{2}$ $(m \geq 1)$
we get combining these inequalities that
	\begin{equation}
    E_0 \, \leq \, E_m 	\qquad  \forall \, m \geq 1
	\end{equation}
if we can find $t > \frac{B}{2}$ such that
	\begin{equation} \label{ent}
    t^2 \left\{ 1+\frac{1}{2}\Big(t^2 - \frac{B^2}{4} \Big) \right\} \delta 
	-2 \eps \Big( t-\frac{B}{2}\Big) < 0
	\end{equation}
where $\delta = 1 + \sqrt{e_o[-2] +1}$

In the estimate for $\delta$ we can use our best variational upper bound.
Inequality \eqref{ent} will be satisfied if $B$ is less than some value $B_0$,
so that in this range $m=0$ is the ground state. In order to see when $m \neq
0$ is a ground state, we use the following trial wave function $\psi(r)$ for a
state with angular momentum $m'$.
	\begin{equation}
	\psi(r) \, = \, r^{m'-m} \, \psi_m(r)  \qquad m' \geq m 
	\end{equation}
where $\psi_m(r)$ is the exact ground state wave function for the state with
angular momentum $m$. An integration by parts shows that
	\begin{gather}
	\int_{0}^{\infty} \ud r \, r \, \big[{\psi_{m}'}^2 \, r^{2(m'-m)} + 2(m'-m) r^{2(m'-m)-1}
	\, \psi_m' \psi_m \big] \nonumber \\
	= \, -\int_0^{\infty} \ud r \, r^{2(m'-m)} \, \psi_m \, (r\psi_m)'
	\end{gather}
Therefore if we use the fact that
	\begin{equation}
	\frac{\eps^2}{r}(r\psi_m')' \, = \, [V_m(r) - e_m]\psi_m
	\end{equation}
We see that
	\begin{align}
	\int_{0}^{\infty}& \ud r \, r \, \big[\eps^2 {\psi'}^2  + V_{m'}(r)\psi^2\big]  =
	e_m  \, \int_0^{\infty} \ud r \, r \, \psi^2  \, \nonumber \\ 
	&+ \, \eps^2 2 m' (m'-m)  \int_{0}^{\infty} \ud r \, r^{2(m'-m)-1} \psi_m^2
	\end{align}
and we conclude that
	\begin{equation}
	e_{m'} \, \leq \, e_m + \eps^2 2 m' (m' - m) 
	\frac{\langle r^{2(m'-m-1)} \rangle_m}{\langle r^{2(m'-m)} \rangle_m}
	\end{equation}
In particular
	\begin{equation}
	e_{1} \, \leq \, e_0 + 2 \eps^2  \frac{1}{\langle r^{2} \rangle_0}
	\end{equation}
If we have a lower bound $c$ on $\langle r^{2} \rangle_0$ then we see that
	\begin{equation}
	E_1 \, < \, E_0 
	\end{equation}
if
	\begin{equation} \label{2ec}
	B \, >  \, \frac{2\eps}{c}
	\end{equation}
We can use for the lower bound $c$  the one given in equation \eqref{rmoy}
	\begin{equation}
	c \, =  \, \frac{\lambda}{2} - \sqrt{e_0[\lambda] + (\frac{\lambda}{2})^2}
	\end{equation}
which is satisfactory when $B$ is not too large, but which becomes negative for
large $B$. We can repair this by using the fact \cite{AHS} that if $f$ is an increasing function of $r$, its expectation
value in the ground state is lowered by adding to the potential a new
increasing potential. We can find a useful comparison potential
	\begin{equation}
	W \, =  \, a_1 r^2 + a_2 r^4 +a_3 r^6	\, \geq \, V
	\end{equation}
which has a ground state wave function of the form
	\begin{equation}
	\psi \, =  \, e^{b_1r^2 - b_2r^4}	\qquad b_2 > 0
	\end{equation}
so that $\langle r^2 \rangle_W$ can be computed explicitly for this potential
and we can take $c = \langle r^2 \rangle_W$  in equation \eqref{2ec}, which
gives a more satisfactory result for large $B$. 
 
In any case we see that the state $m = 1$ if favoured over the state $m=0$  if
$B$ is larger than some value, and by continuity there must exist a field for
which both states have equal energy. But in order to prove that the ground
state is $m= 1$ when $B$ is in some range requires to show that $E_m > E_1$
$\forall m \geq 2$.  For this purpose let us consider $m$ as a continuous
parameter. Then
	\begin{equation} \label{dEm}
	\dpa{E_m}{m} \, = \, 2 \eps^2 m \langle \frac{1}{r^2} \rangle_m -\eps B
	\end{equation}
If we can show that $ \dpa{E_m}{m} \geq 0$ for all $m \geq 1$, then we will have
shown that $E_m > E_1$. When $m \geq 1$ we have
	\begin{equation} \label{cauchy}
	\langle \frac{1}{r^2} \rangle_m \, \geq \, \frac{1}{\langle r^2 \rangle_m}
	\end{equation}
and
	\begin{equation} \label{potmoy}
	\frac{(\eps m)^2}{\langle r^2 \rangle_m} + \langle r^2 \rangle_m^2 +
	\lambda \langle r^2 \rangle_m \, \leq \, e_m
	\end{equation}
In order to get a variational bound on $e_m$ we can use the trial  wave  function
$\psi = r^m e^{-ar^2}$, which gives
	\begin{equation} \label{emp}
	e_m \, \leq \, \frac{m+1}{m+2} V_{m+2} (x_{m+2})
	\end{equation}
where $x_m$ is the value of $x$ which minimises
	\begin{equation} 
	V_m(x) \, = \, \frac{(\eps m)^2}{x} + x^2 +\lambda x
	\end{equation}
Noting that equation \eqref{potmoy} implies that
	\begin{equation}	\label{rmp}
	\langle r^2 \rangle_m \, \leq \, x_m + \sqrt{e_m + V_m(x_m)}
	\end{equation}
one can see by combining equations \eqref{dEm}, \eqref{cauchy}, \eqref{emp}
and \eqref{rmp} that $E_m \geq E_1$ for all $m \geq 2$ if
	\begin{equation} 
	\frac{B^2}{8} \, < \, \frac{1}{1+c^2}
	\end{equation}
with
	\begin{equation} 
	c^2 \, = \, \frac{1}{\eps^2}\, |\lambda| \, \left[x_1 + 
	\sqrt{V_{1+2m}(x_{1+2m}) - V_1(x_1) } \right]^2
	\end{equation}
which implies that $B$ should be less than some value. 

We give in the table \ref{tabineq} some numerical values for the bounds
obtained by these methods. 
	\begin{table}	[!ht] 
	\begin{center}
	\begin{tabular}{|c|c c c|c|c c c|c|c c p{0.0cm}|} 
	\hline
    $\quad \eps \quad$	&  \multicolumn{3}{c|} {$E_0 < E_1$}	&  $B_{0-1}$  &
	 \multicolumn{3}{c|} {$E_1 < E_m$}  &  $B_{1-2}$  &   \multicolumn{3}{c|} {$E_1 < E_0$}   \\
 	\hline\hline																		
     0.01	&  0.0  &-& 0.005  &    0.011  &  0.0  &-&  0.026	  &    0.030	&	0.022  &-& \\
     0.05	&  0.0  &-& 0.024  &    0.054  &  0.0  &-&  0.124	  &    0.163	&	0.146  &-& \\
      0.1	&  0.0  &-& 0.047  &    0.121  &  0.0  &-&  0.221	  &    0.359	&	0.366  &-& \\
      0.2	&  0.0  &-& 0.088  &    0.340  &  0.0  &-&  0.364	  &    0.878	&	1.788  &-& \\
      0.5	&  0.0  &-& 0.189  &    1.610  &  0.0  &-&  0.609	  &    2.745	&	2.834  &-& \\
      1.0	&  0.0  &-& 0.310  &    3.686  &  0.0  &-&  0.826	  &    5.846	&	4.277  &-& \\
      2.0	&  0.0  &-& 0.469  &    7.816  &  0.0  &-&  1.066	  &   11.910	&	8.141  &-& \\
  	\hline
	\end{tabular}
	\caption{Results of the inequalities on the energies and values $B_m$}
	\label{tabineq}
	\end{center}
	\end{table}	
They show that whereas the range of values of $B$ for which $E_0 < E_1$ and
$E_1 < E_0$ is reasonably estimated for $\eps \lesssim 0.1$, there is no range
of values of $B$ for which our bounds show that $m= 1$ is the  ground state
except when $\eps$ is very small (0.01) But in this range WKB works perfectly
well.  Obviously we have too poorly estimated the effect of the kinetic energy
and that of the centrifugal barrier. Numerical computations for example show
that the replacement of   $\langle \frac{1}{r^2} \rangle_1$ by
$\frac{1}{\langle r^2 \rangle_1}$  is not appropriate when $\eps$ or $B$ are
too large. 

In conclusion, even in two dimensions improved rigorous bounds on the critical
values of the magnetic field are needed, and the WKB method for which we have
no estimate of the error gives the best analytic results.


\section{conclusion}

It could be of course quite interesting to see an experimental verification of
these surprising effects of the magnetic field. Even though we have found them
in the case of a double-well, we think that the details of the potential do not
matter too much. What is needed is a potential whose minimum is taken
sufficiently far from the origin. 

We have thought of two possible fields where one could observe such effects. 
The first one is molecular physics where often the dynamics of electrons or
protons is modelled by the motion of a quantum particle in a double-well
(although admittedly often a one-dimensional one.) If we consider the case of
the electron in the rotationally symmetric double-well, the smallest value of
the critical field where the $m = 1$ and $m = 0$ states are degenerate is about
$15$ Tesla if we take for the depth of the potential $1$ eV and for the
distance to the origin of the minimum 2 $\AA$. For protons the situation is
more favourable since a field of $5$ Tesla can create a degeneracy when the
depth is kept to $1$ eV and the minimum is at a distance of $1.5$ $\AA$.
Obviously a more detailed investigation is needed if one wants to see these
unusual effects (like a change from diamagnietism to paramagnetism) in
molecules.

The other field is that of Bose-Einstein condensates of very
cold atoms, which recently has made spectacular progress. 
If we consider free charged bosons in a magnetic field and in a potential
$V(\frac{r}{r_0})$ one can show that there is a Bose-Einstein condensation
in the ground state in three dimensions, in the limit 
$r_0$ going to infinity,
for all potentials which have a quadratic dependence of $r$ near the origin.
Our result supports therefore that free charged bosons in their condensate 
would show a phase transition when one varies the magnetic field. This
transition would manifest itself by jumps of the magnetisation at some
critical values of the magnetic field. The phenomenon would probably persist
in a dilute gas of charged bosons in a neutralising background. It is however
probably quite difficult to create such a jellium in the laboratory and this
remains a challenging task.

\section{Acknowledgements}

We thank Ph. Martin and N. Datta for some useful discussions on the 
Bose-Einstein condensation in the presence of a magnetic field.


\end{document}